\documentclass{IEEEtran}
\usepackage[latin1]{inputenc}
\usepackage{times,amsmath}
\usepackage{amsfonts}
\usepackage{pstool}
\usepackage{subcaption}
\usepackage{multirow}
\usepackage{multicol}
\usepackage{enumerate}
\usepackage{graphicx}
\usepackage{MnSymbol}
\usepackage{stfloats}
\usepackage[table]{xcolor}
\usepackage[square, comma, sort&compress, numbers]{natbib}
\usepackage{nohyperref}
\usepackage{algorithm,algorithmic}
\usepackage{bigints}
\usepackage{amsmath}
\usepackage[justification=centering]{caption}

\newcounter{tempEquationCounter}
\newcounter{thisEquationNumber}

\makeatletter
\newcommand{\vast}{\bBigg@{4}}
\newcommand{\Vast}{\bBigg@{5}}
\newcommand\numeq[1]%
  {\stackrel{\scriptscriptstyle(\mkern-1.5mu#1\mkern-1.5mu)}{=}}
\makeatother

\graphicspath{ {Figures/} }

\begin{document}

\title{On the Potential of Re-configurable Intelligent Surface (RIS)-assisted Physical Layer Authentication (PLA)}
\author{
\IEEEauthorblockN{Hala Amin, Waqas\ Aman, Saif Al-Kuwari, 
\\
Division of Information and Computing Technology, College of Science and Engineering, \\Hamad Bin Khalifa University, Qatar Foundation, Doha, Qatar. 
 \\
haam51711@hbku.edu.qa, waman@hbku.edu.qa, smalkuwari@hbku.edu.qa
}
}

\maketitle

\maketitle

\begin{abstract}
    Re-configurable Intelligent Surfaces (RIS) technology is increasingly becoming a potential component for next-generation wireless networks, offering enhanced performance in terms of throughput, spectral, and energy efficiency.  However, the broadcast nature of RIS-assisted wireless communication makes it vulnerable to malicious attacks at the physical layer. At the same time, physical layer authentication is gaining popularity as a solution to secure wireless network, thwarting different attacks such as cloning, spoofing, and impersonation by using the random features of the physical layer.
    In this paper, we investigate RIS-assisted wireless communication systems to unlock the potential of using RIS for physical layer authentication (PLA). In particular, we exploit two distinct features of the physical layer: pathloss and channel impulse response (CIR) for PLA in RIS-assisted wireless communication. We construct hypothesis tests for the estimated features and derive closed-form error expressions. Further, we consider the critical error, i.e., missed detection, as our objective function to minimize by optimizing the phase shift of the RIS pannel. We compare the performance of our proposed mechanisms with PLA schemes using the same features but with no RIS.  Furthermore, we thoroughly evaluate our proposed schemes using performance metrics such as the probability of false alarm (PFA), the probability of missed detection (PMD), and the receiver operating characteristic (ROC) curves. The results demonstrate clear positive impact of RIS on PLA, as it effectively reduces PMD values to zero when determining the optimal phase shift.
\end{abstract}


\section{Introduction}
The need for innovative solutions has increased with the increasing use of wireless devices for advanced communication in cellular networks. These solutions aim to boost the efficiency of energy and spectrum utilization while concurrently elevating the reliability and security of wireless communication systems. Researchers and engineers are evaluating and proposing new approaches to achieving these evolving goals in this era of newfound technological possibilities \cite{wang2023road}.
Ambitious goals set for the fifth-generation (5G) wireless network, including the three core services, enhanced mobile broadband (eMBB), ultra-reliable and low-latency communications (URLLC), and massive machine-type communications (mMTC), have been substantially realized \cite{8476595,guo2022customized}. This achievement is mainly attributed to pivotal enabling technologies such as ultra-dense networks (UDN), massive multiple input, multiple output (MIMO), and millimeter wave (mmWave) communication \cite{kulkarni2021key,gao2015mmwave,8387217}. 

The upcoming sixth-generation (6G) cellular networks are expected to facilitate a more comprehensive range of applications and services than 5G wireless networks \cite{jiang2021road, dang2020should}. The evolutionary objectives of 6G frameworks bring forward transformative elements like data-centric, immediate, immensely scalable, and omnipresent wireless connectivity coupled with integrated intelligence \cite{8808168}. Therefore, it remains crucial to prioritize research efforts toward developing innovative, spectral, energy-efficient, secure, and economically viable solutions for wireless networks after the 5G era. 

\begin{figure}[htb!]
    \centering
    \includegraphics[width=1\linewidth]{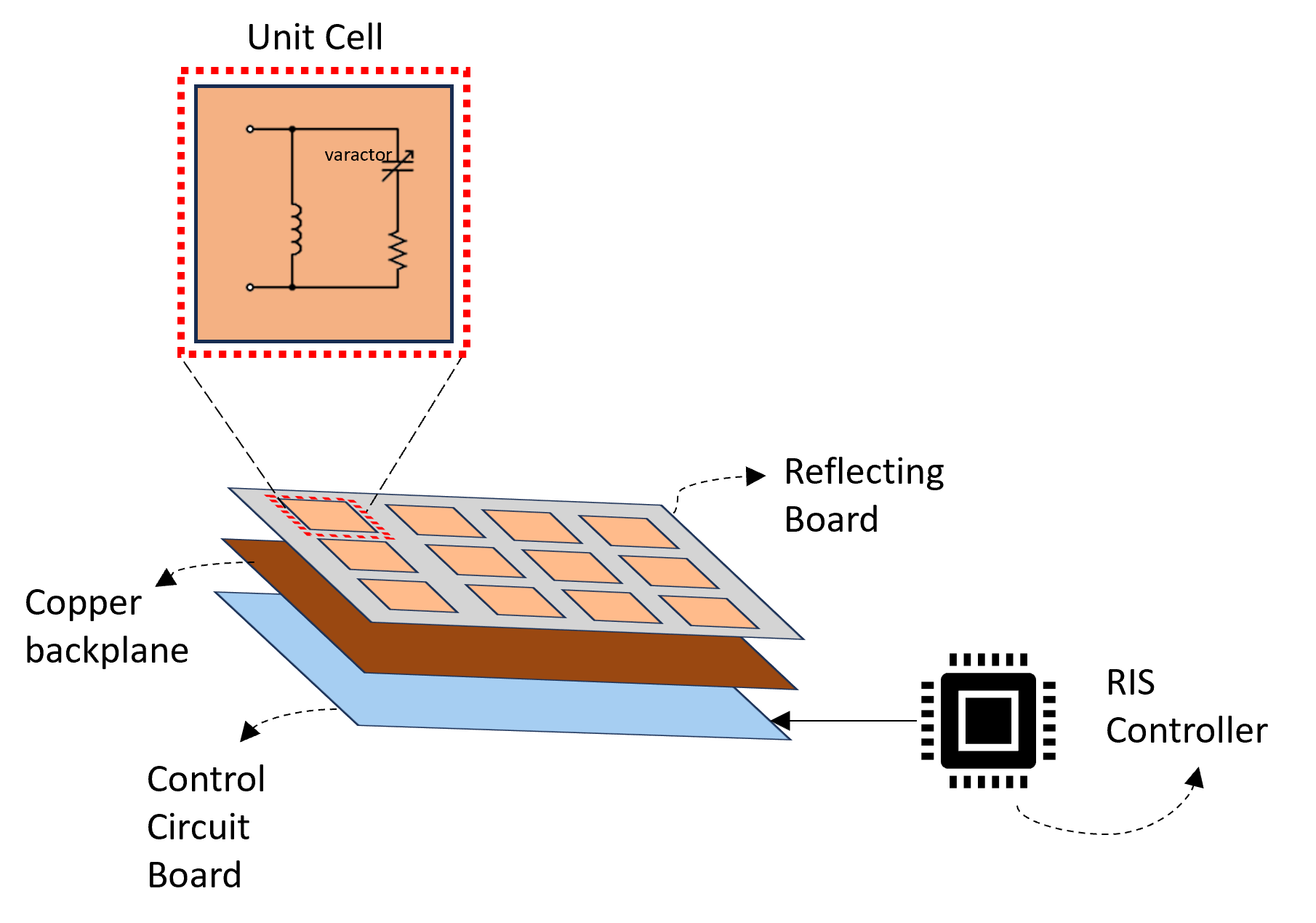}
    \caption{RIS architecture composed of three layers along with a controller}
    \label{fig:RIS Arch}
\end{figure}

Re-configurable Intelligent Surface (RIS) is considered one of the key enablers for 6G and beyond wireless communication systems due to its enhanced performance over no-RIS/traditional wireless communication systems \cite{8796365}. Typically, RIS comprises tiled metamaterial elements designed to reflect incoming radio signals with controllable phases using a diode array as illustrated in Figure \ref{fig:RIS Arch}. By intelligently controlling the phase and amplitude of reflected signals, RIS can shape the propagation environment, enhance signal strength, mitigate interference, and extend coverage range. This capability holds profound implications for the design and optimization of future communication systems \cite{liu2021reconfigurable}.  In the context of 6G, RIS offers several key advantages that align with the evolving requirements of next-generation networks including coverage and connectivity \cite{10315212}, security \cite{9504435}, throughput, energy, and spectral efficiency \cite{8796365,aman2021effective}. 

 The inherent broadcast nature of wireless channels in RIS-assisted systems poses a significant challenge in establishing a secure communication environment \cite{porambage2021roadmap}.  Consequently, there is a pressing need for a dedicated and effective method to ensure wireless communication security. Physical Layer Security (PLS) emerges as a promising solution to overcome the limitations of upper-layer security protocols. Although traditional cryptography-based methods offer adequate security for such systems, they are dependent on secret keys and introduce unavoidable computational overhead \cite{illi2023physical}. On the other hand, PLS leverages the unique statistical characteristics of physical channels that can enhance both the confidentiality and authentication aspects of wireless systems while maintaining an acceptable energy consumption that is ideal for energy-constrained devices \cite{sanenga2020overview, jorswieck2015broadcasting}. A key component of PLS is Physical Layer Authentication (PLA), which serves to authenticate legitimate nodes and thwart impersonation, cloning, and spoofing attempts \cite{PLA:Survey:2021}. Generally, PLA exploits a single or multiple features of the physical layer for authentication purposes. To date, a variety of fingerprints are exploited for PLA which includes, Carrier offsets, I/Q imbalance, CFR, CIR, pathloss \cite{aman2023security}, and device physical location \cite{10323824} for different terrestrial and non-terrestrial wireless communication systems.
  RIS has been rigorously studied for improved throughput, coverage, spectral, and energy efficiency. On the other hand, there are significant studies reported on RIS for PLS mainly focusing on the confidentiality or secrecy performance of the system with different assumptions and system models \cite{yang2020secrecy,9740145,kaveh2023secrecy,zhang2023robust}.



\subsection{Related Work}

To the best of our knowledge, the discussion on the use of RIS for PLA begins in \cite{9982485} where the authors proposed challenge-response (CR) mechanisms in the context of PLA. The mechanism focuses on authentication through changes in the electromagnetic environment by partially controlling the wireless channel. This involves introducing an RIS with modified changes in its configuration during each transmission. The authors discussed in broad terms the role of RIS in partially controlling the wireless channel for PLA. However, they did not provide a system model of mathematical expressions that details the technique necessary to achieve this objective.

Next, very recently, in \cite{10199455}, the authors suggested an innovative design for a RIS, introducing the concept of a Hybrid RIS (H-RIS). The H-RIS retains the standard reflective properties of the conventional RIS configurations incorporated with additional functionality. This reflects the incident signal while selectively absorbing a portion of it to facilitate a collaborative authentication process with the receiver to perform channel estimation. Meanwhile, the remaining portion of the signal is effectively directed by H-RIS toward its intended destination through the conventional reflective capabilities of the H-RIS.
The outlined authentication method employs an active RIS that acts as a relay. However, this introduces additional signal processing requirements that increase costs and computational overhead compared to the passive RIS alternative. 
The authors evaluated their scheme by analyzing how varying the number of elements in the RIS impacts PMD while maintaining a constant Signal-to-Noise Ratio (SNR) value. This paper does not provide any insights on the design of phase shift or the role of phase shift in PLA. 
\subsection{Contribution}
In this paper, for the first time, we systematically study PLA in RIS-assisted wireless communication. Specifically, we demonstrate the impact of using RIS on the performance of PLA in wireless communications.  Our contributions can be summarized as follows: 
\begin{itemize}
    \item We systematically exploit two distinct physical layer features: path-loss and CIR for PLA in RIS-assisted wireless communication. For pathloss, we construct the binary hypothesis testing and derive the closed-form expressions for the two inherent errors, i.e. false alarm and missed detection. For CIR, we split the complex CIR into magnitude and phase components and exploit them individually for PLA. We construct the binary hypothesis testing and derive the error expressions.
    \item We formulate optimization programs to solve the phase shift problem to minimize the missed detection probability for path loss and CIR. We solve the optimization program through an exhaustive search method. 
    \item We compare the performance of our proposed mechanism against the baseline schemes which use the same fingerprints but with no-RIS setup. To the best of our knowledge, there is no work in the literature that compares the performance of non-RIS wireless systems and RIS-assisted systems in terms of PLA. 
    \item We also perform a Monte-Carlo simulation to validate the analysis. We observe a perfect match of the Monte-Carlo and analytical results. We use a Matlab-based simulator named \textit{SimRis} for generating channel realizations.
    
    \item We assess the effectiveness of PLA schemes for both non-RIS and RIS-assisted communication systems, using metrics such as PFA, PMD, and ROC. Our analysis highlights the superior performance of the RIS-assisted PLA algorithm compared to its non-RIS counterpart.
\end{itemize}
\subsection{Organization}
The rest of this paper is organized as follows:
In Section \ref{s: s2} we describe our system model. Section \ref{s: s3} introduces our RIS-assisted PLA schemes. We discuss the results of the evaluation and simulation of the proposed techniques in Section \ref{s: s4}. We conclude the paper in Section \ref{s: s5}, summarizing our findings and suggesting some future work.

\section{System Model}
\label{s: s2}
\begin{figure}
    \centering
    \includegraphics[width=0.9\linewidth]{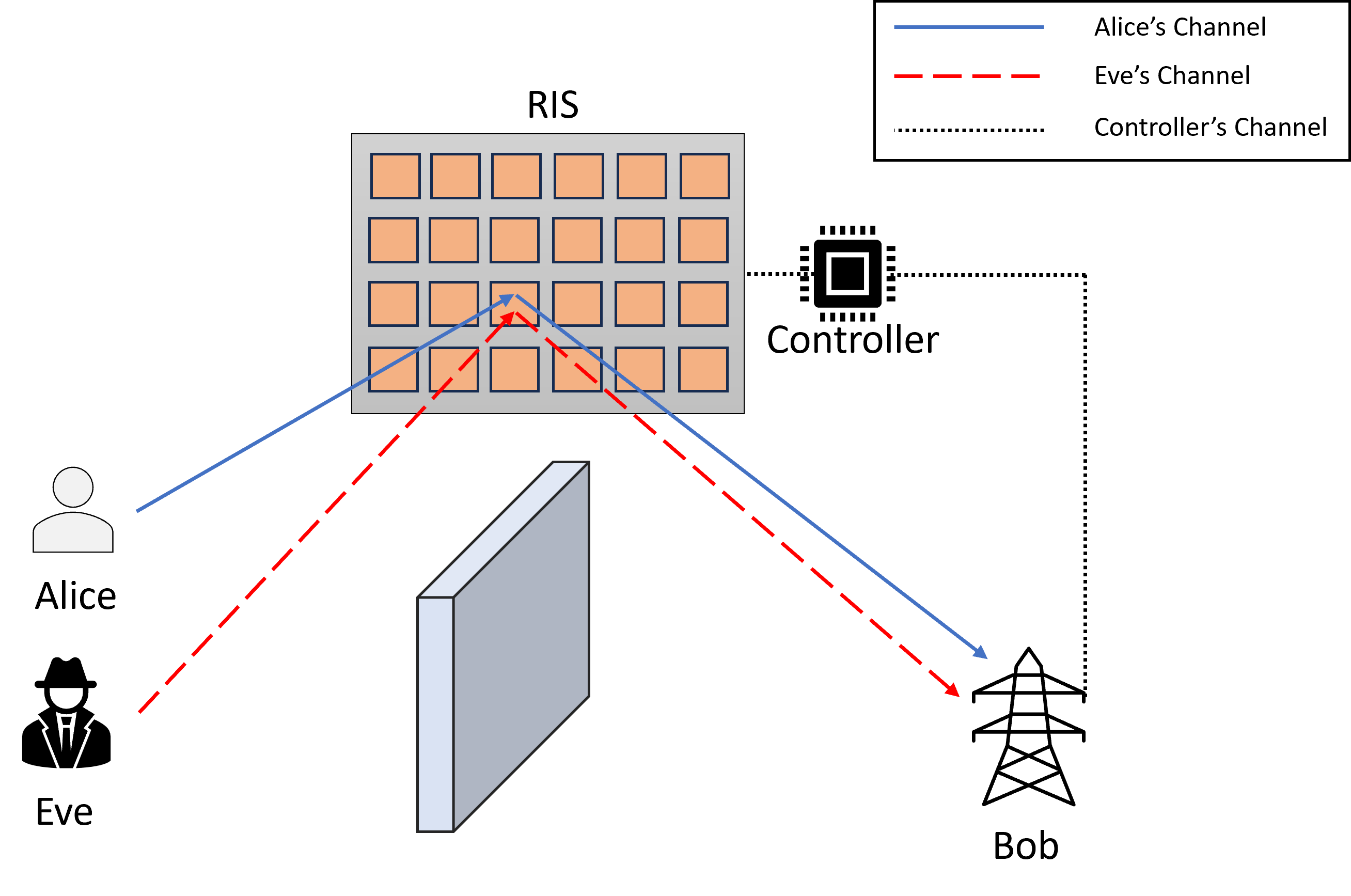}
    \caption{RIS-assisted Wireless Communication System}
    \label{fig:RIS wireless system}
\end{figure}

As illustrated in Figure \ref{fig:RIS wireless system}, our system features \textit{Alice} as the legitimate node and \textit{Eve} as the malicious node, which acts as transmitters. Meanwhile, \textit{Bob} serves as a receiver and performs the authentication process. Alice and Eve's communication with Bob is assisted by the \textit{RIS}, as the direct transmission path between the transmitters and the receiver is blocked. Bob acts as the base station and controls the RIS to adjust its phase shifts for optimal signal redirection. 

\begin{figure}
    \centering
    \includegraphics[width=0.65\linewidth]{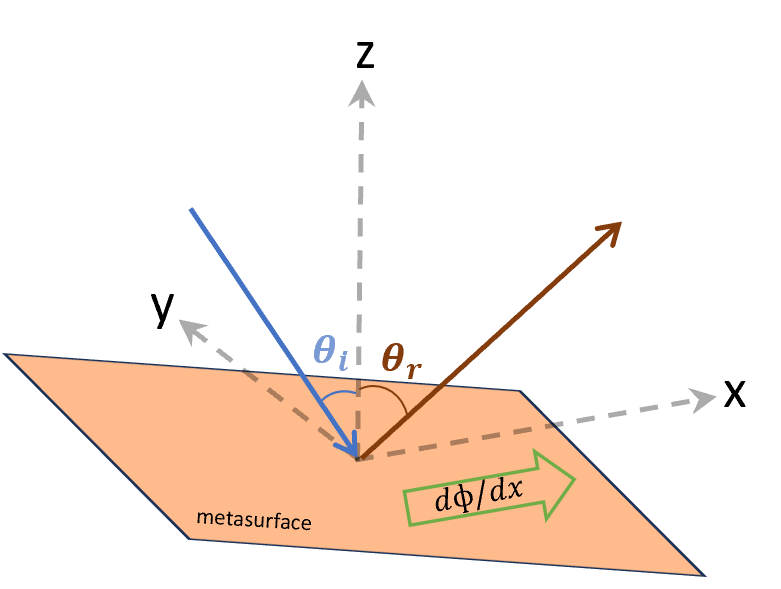}
    \caption{An illustration of 2D RIS element that reflects the impinging ray to an arbitrary direction}
    \label{fig:RISelement}
\end{figure}

RIS comprises metasurfaces that reflect the impinging signals in the desired direction by manipulating the metasurfaces' phase shifts or orientation. Figure \ref{fig:RISelement} illustrates the metasurface design and illustrates its functionality. 
The angle of reflection $\theta_r \in [0,2\pi]$ is a function of the angle of incident $\theta_i$ and the change in the slope of the element w.r.t a given axis. Thus, the angle of reflection for a particular RIS element can be expressed as \cite{liu2021reconfigurable}:
\begin{equation}
\label{eq: theta_r}
    \theta_r = \sin^{-1}{ \left[ \sin{\theta_i} + \frac{\lambda}{2\pi n_1} \frac{d\Phi}{dx} \right]}
\end{equation}
where $\lambda$ is the wavelength of the transmitted signal, $n_1$ is the refractive index of the air medium, $\frac{d\Phi}{dx}$ is the gradient of phase discontinuity, which is the rate of change of phase $\Phi$ w.r.t. the spatial variable $x$. In other words, it means that the angle of reflection can be controlled through the change in the slope of the metasurface element. Note that this change is governed by the controller of the RIS which can be remotely controlled by the base station/Bob. We assume that Eve is smart enough to sense the communication medium and transmit its malicious information to Bob once finds it idle. 

\section{Proposed PLA Mechanisms in RIS-assisted Wireless Communication}
\label{s: s3}
In this section, we propose two RIS-assisted PLA methods each based on a physical layer characteristic, i.e. pathloss and CIR. We explore each property in RIS-assisted wireless communication systems and their employment to achieve the desired PLA algorithm.

\subsection{Pathloss based PLA}
We use a pathloss model for RIS-assisted wireless communications grounded in the principles of physics and the electromagnetic properties inherent to RIS'. 
Using the far-field assumption and the fact that the pathloss of a RIS comprising $N$ elements is almost the same as the path loss of a single RIS element, the pathloss is given as \cite{ozdogan2019intelligent}:.
\begin{equation}
\label{eq.Pathloss-RIS}
\text{PL} = \frac{G_t G_r}{(4\pi)^2}\left(\frac{a b}{d_ir}\right)^2 \times \cos^2(\theta_i) \left( \frac{\sin(\frac{\pi b}{\lambda}(\sin(\theta_i) - \sin(\theta_r)))}{\frac{\pi b}{\lambda}( \sin(\theta_i) - \sin(\theta_r))} \right)^2
\end{equation}
where $a\times b$ is the dimension of the 2D RIS element, $G_t$ and $G_r$ are the antenna gains of the transmitter and receiver, respectively. The distance between a transmitter and the RIS is denoted by $d_i$, while $r$ is the distance between the RIS and the receiver. 
However, the estimated pathloss at the Bob is noisy due to noise present in the received signal, therefore the noisy estimate of the pathloss can be expressed as:
\begin{equation}
\label{eq.estimated PL}
\hat{\text{PL}} = \text{PL} + n,
\end{equation}
where $\text{PL}$ is the actual pathloss defined in Eq. \ref{eq.Pathloss-RIS}, and $n \sim \mathcal{N}(0,\sigma^2)$ is the uncertainty added due to noise from the channel.


\subsubsection{Binary Hypothesis Testing}
Now, the estimated pathloss at Bob might belong to Alice or Eve. Therefore, we define two hypotheses as: 
 \begin{equation}
\label{eq.Hcases}
\begin{aligned}
&\mathcal{H}_0 : \hat{\text{PL}} = \text{PL}_A + n\\
&\mathcal{H}_1 : \hat{\text{PL}} = \text{PL}_E + n
\end{aligned}
\end{equation}
where $\mathcal{H}_0$ represents the null hypothesis, indicating that Alice sent a legitimate signal. However, $\mathcal{H}_1$ represents the alternative hypothesis, indicating that Eve sent a malicious signal. 

In light of the presented hypotheses, we now proceed to introduce the test statistics, a pivotal analytical tool that will enable us to discern the relative likelihood of each hypothesis. The test statistics (TS) is defined as:
\begin{equation}
\label{eq.teststat}
\text{TS} = |\hat{\text{PL}} - \text{PL}_A|,
\end{equation}
where $\text{PL}_A$ is the ground truth or actual pathloss of the legitimate node obtained earlier by Bob. Using the above TS, the binary hypothesis testing could be expressed as:
\begin{equation}
	\label{eq:H0H1-PL-RIS}
	 \begin{cases} \mathcal{H}_0 (\text{Alice}): & \text{TS}= |\hat{\text{PL}} - \text{PL}_A| < \epsilon \\ 
                  \mathcal{H}_1 (\text{Eve}): & \text{TS}= |\hat{\text{PL}} - \text{PL}_A| > \epsilon \end{cases},
\end{equation}
where $\epsilon$ is a predefined parameter serving as a threshold. The binary hypothesis test may alternatively be characterized as:
\begin{equation} 
\label{eq:bht}
\text{TS} \gtrless_{\mathcal{H}_1}^{\mathcal{H}_0} {\epsilon}.
\end{equation}

\subsubsection{Performance Analysis}

\label{error-RIS}
To assess the performance of the proposed PLA mechanism, we compute two important probabilities: PFA and PMD. The PFA can be expressed as:

\begin{equation}
    \text{P}_{fa} = \text{Pr}(\text{TS}|\mathcal{H}_0 > \epsilon).
    \label{eq:pfa-RIS}
\end{equation}
The PFA equals finding the probability of the event that TS given that the transmitter was Alice is greater than the threshold $\epsilon$. Under the null hypothesis,  TS$|\mathcal{H}_0 = |\text{PL}_A+n-\text{PL}_A| = |n| > \epsilon_{th}$. One can see that TS$|\mathcal{H}_0 $ is now a folded normal distributed random variable (RV). So, the probability is now the  complementary cumulative distribution function (CCDF) of the folded normal RV TS$|\mathcal{H}_0 $ which can be expressed as:
 \begin{align}
 \label{eq:pfaccdf-RIS}
    \text{P}_{fa} & =1- \frac{1}{2} \text{erf}(\frac{\epsilon + \mu_{\text{n}}}{\sqrt{2}\sigma})+\text{erf}(\frac{\epsilon-\mu_{\text{n}}}{\sqrt{2}\sigma}) \stackrel{(a)}=2Q(\frac{\epsilon}{\sigma}), 
 \end{align}
 where 
 $Q(x) = \frac{1}{\sqrt{2\pi}} \int_{x}^{\infty} e^{-\frac{t^2}{2}} dt$ is a standard $Q$-function and $\text{erf}(x)$ is the error function, $\mu_n$ is the mean of RV $n$. Eq.\ref{eq:pfaccdf-RIS} (a) is a result of the use of standard relation between $Q$ and error functions with $mu_{\text{n}}=0$. Now, for a given $\text{P}_{fa}$, we can select an appropriate value for $\epsilon$ in a way that minimizes the $P_{md}$. The threshold $\epsilon$, could be computed from Eq.\ref{eq:pfaccdf-RIS} as per Neyman-Pearson lemma and is given as:
\begin{equation}
\label{eq:threshold}
    \epsilon = \sigma Q^{-1}(\frac{P_{fa}}{2}),
\end{equation}
where $Q^{-1}(.)$ is the inverse Q-function. Next, we compute PMD which can be expressed as:

\begin{equation}
    \text{P}_{md} = \text{Pr}(\text{TS}|\mathcal{H}_1 \leq \epsilon).
    \label{eq:pmd-RIS}
\end{equation}
So it means PMD equals finding the probability of the event that $\text{TS}|\mathcal{H}_1$ takes values less than $\epsilon$. One can find that TS$|\mathcal{H}_1 \sim \mathcal{FN} (\mu_{\text{TS}|\mathcal{H}_1}, \sigma_{\text{TS}|\mathcal{H}_1}^2)$ with $\mu_{\text{TS}|\mathcal{H}_1}=\exp{(\frac{-(\text{PL}_E - \text{PL}_A)^2}{2\sigma^2})}\sigma\sqrt{\frac{2}{\pi}}+(\text{PL}_E - \text{PL}_A)(1-2\text{CDF}_n(-\frac{\text{PL}_E - \text{PL}_A)}{\sigma})$, $\sigma_{\text{TS}|\mathcal{H}_1}^2=(\text{PL}_E - \text{PL}_A)^2+\sigma^2-\mu_{\text{TS}|\mathcal{H}_1}^2$ . The $P_{md}$ is basically the CDF of $\text{TS}|\mathcal{H}_1$ which is given as:

\begin{equation}
\label{eq:pmdcdf-RIS}
\begin{aligned}
\text{P}_{md} = {\frac  {1}{2}}\left[{\mbox{erf}}\left({\frac  {\epsilon+\text{PL}_E - \text{PL}_A }{\sigma {\sqrt  {2}}}}\right)+{\mbox{erf}}\left({\frac  {\epsilon- \text{PL}_E + \text{PL}_A }{\sigma {\sqrt  {2}}}}\right)\right].
\end{aligned}
\end{equation}

\subsubsection{Optimal Phase Shift Design}
\label{s: phase shift design}

The efficacy of RIS in the authentication process hinges on its ability to leverage its phase shift. Choosing the optimal phase shift would assist in providing distinguishable transmitter channels and Alice's and Eve's fingerprints. One can see that the above PMD expression is a function of phase shift or $\frac{d\Phi}{dx}$of the RIS. We can write it as $\text{P}_{md}(\phi)$ where $\phi=\frac{d\Phi}{dx}$.
We find the optimal phase shift at the minimal $\text{P}_{md}$ values and the relevant optimization problem is expressed as:
\begin{align}
  &  \text{arg} \min_{\phi} \\  &{\frac  {1}{2}}\left[{\mbox{erf}}\left({\frac  {\epsilon+\text{PL}_E(\phi) - \text{PL}_A(\phi) }{\sigma {\sqrt  {2}}}}\right)+{\mbox{erf}}\left({\frac  {\epsilon- \text{PL}_E(\phi) + \text{PL}_A(\phi) }{\sigma {\sqrt  {2}}}}\right)\right]. \nonumber 
\end{align}
We use an exhaustive search method to find the optimal value for the phase shift $\phi$. 

\subsection{CIR based PLA}
\label{CIR-RIS}

The second mechanism is based on the use of CIR to achieve authentication at the physical layer of RIS-assisted wireless communication. 


 The received signal $y$ in the baseband at Bob when Alice transmits a symbol $x$  is:

\begin{equation}
\label{eq.CIR-RIS-Alice}
 y_{A} = \mathbf{h}_A^* \,\mathbf{\Phi}\, \mathbf{g}_A \,x + n, 
\end{equation}
where $\mathbf{\Phi} = diag{[e^{j\psi_1} \:e^{j\psi_2} \:e^{j\psi_3}\ldots e^{j\psi_N}]}$ is a diagonal matrix of dimension $N\times N$ that contains the phase shifts of the RIS elements, $\mathbf{h}_A = [h_1\ldots h_N] \in \mathcal{C}^{N\times 1}$ is the channel vector from Alice to RIS with  $h_{n^{th}}\sim \mathcal{CN}(0, 1)$ is the channel gain from Alice to the $n^{th}$ RIS element , $\mathbf{g}_A = [g_1\ldots g_N] \in \mathcal{C}^{N\times 1}$ is the channel vector from RIS to Bob with $g_{n^{th}}\sim \mathcal{CN}(0, \sigma^2)$ is the channel gain from the $n^{th}$ RIS element to Bob when Alice transmits, and $n \sim \mathcal{CN}(0, \sigma^2)$ is the noise.
Conversely, the transmitted symbol $x$ by Eve will be received on Bob side represented as follows:
\begin{equation}
\label{eq.CIR-RIS-Eve}
 y_{E} = \mathbf{h}_E^* \, \mathbf{\Phi}\, \mathbf{g}_E \,x + n, 
\end{equation}
where $\mathbf{h}_E$ is the channel vector from Eve to RIS and $\mathbf{g}_E$ is the channel vector from RIS to Bob when Eve transmits.
Now, the baseband noisy CIR  by an anonymous transmitter can be expressed as
\begin{equation}
\label{eq.CIR-RIS}
 \zeta = \mathbf{h}^*\,\mathbf{\Phi}\, \mathbf{g} + n 
\end{equation}

As the obtained CIR in the baseband is a complex number function, we write it in its magnitude and phase representation. The magnitude and [hase] of the estimated CIR can be expressed as:
\begin{align}
    \zeta^{m}=|\zeta| = \sqrt{\text{Re}(\zeta)^2+\text{Im}(\zeta)^2},    \zeta^{p}= \angle \zeta   =\text{Tan}^{-1}\left(\frac{\text{Im}(\zeta)}{\text{Re}(\zeta)}\right),
\end{align}
where $\text{Re(.)}$ and $\text{Im}(.)$ denotes the real and imaginary part of a complex number, and $\text{Tan}^{-1}$ is the inverse tangent function.
\subsubsection{Binary Hypothesis Testing}
\label{BHT-CIR-R}

In order to discern between legitimate and malicious nodes, we use a binary hypothesis test. As the CIR van be split into magnitude and phase components, therefore, instead of a single binary hypothesis test we have two binary hypothesis tests. 
%


We define the test statistics for both components as follows:
\begin{equation}
\label{eq:mTS-CIR-RIS}
\text{TS}^m= |\zeta - \mathbf{h}_A^* \,\mathbf{\Phi}\, \mathbf{g}_A  |
\end{equation}
\begin{equation}
\label{eq:pTS-CIR-RIS}
\text{TS}^p= |\zeta^p -\angle ( \mathbf{h}_A^* \,\mathbf{\Phi}\, \mathbf{g}_A )|. \nonumber
\end{equation}

Similarly, we now have two distinct hypothesis tests: one indicates the intrinsic design (i.e., subtract the ground truth and take the magnitude or modulus of the complex number)  and the other for phase, which is given below:
\begin{equation}
	\label{eq:mH0H1-CIR-RIS}
	\text{BHT}^m \implies \begin{cases} \mathcal{H}_0 (\text{Alice}): & \text{TS}^m= |\zeta - \mathbf{h}_A^* \,\mathbf{\Phi}\, \mathbf{g}_A  | < \epsilon \\ 
                  \mathcal{H}_1 (\text{Eve}): & \text{TS}^m= |\zeta - \mathbf{h}_A^* \,\mathbf{\Phi}\, \mathbf{g}_A  | > \epsilon \end{cases},
\end{equation}
\begin{equation}
	\label{eq:pH0H1-CIR-RIS}
	\text{BHT}^p \implies \begin{cases} \mathcal{H}_0 (\text{Alice}): & \text{TS}^p= |\zeta^p - \angle ( \mathbf{h}_A^* \,\mathbf{\Phi}\, \mathbf{g}_A )| < \epsilon \\ 
                  \mathcal{H}_1 (\text{Eve}): & \text{TS}^p= |\zeta^p - \angle ( \mathbf{h}_A^* \,\mathbf{\Phi}\, \mathbf{g}_A )| > \epsilon \end{cases},
\end{equation}


Note that the reason for proposing two BHTs in the case of CIR is that we noticed the phase shift of the RIS is not changing any of the errors due to the intrinsic design of the test statistics of BHT$^m$, so to incorporate the role of the RIS-phase shift we propose the BHT$^p$.
\subsubsection{Performance Analysis}
\label{error-CIR-R}

The earlier defined error probabilities (PFA and PMD) are chosen to evaluate the proposed PLA mechanism. The PFA for the BHT$^m$ and BHT$^p$ can be written as:
\begin{equation} 
\label{eq:mTCIR_pfa_RIS}
{\text{P}_{fa}}^m = \text{Pr}(\text{TS}^m\mid \mathcal{H}_0 > \epsilon), \ \text{P}_{fa}^p = \text{Pr}(\text{TS}^p\mid \mathcal{H}_0 > \epsilon) 
\end{equation}
One can find that $\text{TS}^m\mid \mathcal{H}_0$ is the absolute of complex Gaussian RV having zero mean and variance $\sigma^2$. So it implies that $\text{TS}^m\mid \mathcal{H}_0$ is a Rayleigh distributed RV with parameter $\sigma$. This leads us to the closed-form expression for ${P_{fa}}^m$, which is given below:
\begin{align}
\label{eq:mTCIR_pfa_RIS}
{\text{P}_{fa}}^m = \exp{(-\frac{\epsilon^2}{2\sigma^2})}.
\end{align}

Next, we observe that finding the distribution of $\text{TS}^p\mid \mathcal{H}_0$ is involved due to the absence of clarity in the literature for that function of RV, so we numerically compute the ${\text{P}_{fa}}^p$ in simulations. 


The PMDs for the magnitude and phase according to the definition can be expressed as:
\begin{equation} 
\label{eq:mTCIR_pmd_RIS}
{\text{P}_{md}}^m = \text{Pr}(\text{TS}^m\mid \mathcal{H}_1 \leq ), \ {\text{P}_{md}}^p = \text{Pr}(\text{TS}^p\mid \mathcal{H}_1 \leq ) 
\end{equation}

Here, finding the nature of both RVs is quite complex, so we numerically compute both errors.
\subsubsection{Optimal Phase Shift Design}
\label{op-CIR}
One can find that  $\text{P}_{md}^p$ is a function of the phase shift of RIS, therefore, one can find the optimal value for the phase shift of RIS $\Phi$ that minimizes the  $\text{P}_{md}^p$. The formulated optimization program is given as:

\begin{align}
    \arg \min_{\psi_i, \forall i \in \{ 1,...N\}\in \Phi}  & \text{Pr}(\text{TS}^p\mid \mathcal{H}_1(\Phi) \leq \epsilon) \\
   & \text{s.t.} \Phi \in \mathcal{F}, \nonumber
\end{align}
where $\mathcal{F}$ is a set containing all possible integer combinations for total $N$ phase shifts. The constraint is due to the fact that in a realistic setup, the phase shifts can be discrete. The optimization problem is a combinatorial optimization problem, as we don't have closed-form expressions for the missed detection probability. Therefore, we solve this problem via an exhaustive search method where for each integer combination of phase shifts we numerically compute the missed detection probability.

\section{Simulation}
\label{s: s4}

\subsection{Setup}
\label{s: parameters}
In order to evaluate our PLA schemes based on simulations, we use MATLAB environment. Our model operates with stationary nodes, strategically placing both legitimate and malicious nodes close to increase the chances of missed detection. The simulation parameters are presented in Table \ref{tab: pathloss parameters}.

\begin{table}[h!]
\centering
\caption{Simulation Parameters}
\begin{tabular}{| c | c |}
\hline
Parameter & Value  \\
\hline\hline
RIS element's length and width ($a$x$b$)  & $0.5m$ x $0.5m$ \\
\hline
Number of RIS elements  & 256 \\
\hline
Legitimate node position & [100,100,1]  \\ \hline
Malicious node position & [90,100,1]  \\ \hline
RIS Position  & [90,90,1]   \\ \hline
Frequency  & $28 \times 10^9 \mathrm{Hz}$  \\ \hline
Transmission power & $1 \mathrm{W}$\\ \hline
Transmitter Antenna Gain & $1000$\\ \hline
Receiver Antenna Gain & $1000$\\

\hline
\end{tabular}
\label{tab: pathloss parameters}
\end{table}

To ensure a realistic simulation of our PLA based on the CIR evaluation, we use the \textit{SimRIS Channel Simulator} \cite{basar2020simris,basar2020SimRIS_2,basar2020SimRIS_3}, which is an open-source MATLAB-based tool that allows simulating channels for RIS-assisted systems in indoor and outdoor environments. The simulator allows to change the surroundings, RIS element count, terminal placements, and operation frequency. Channels at 28 GHz and 73 GHz frequencies can be generated for RIS-enabled communication systems in Indoor Office (InH) and Street Canyon (UMi) settings.

We integrate a scheme drawn from existing literature as a baseline, focusing on achieving PLA in a traditional/non-RIS wireless communication setup, where transmitters communicate directly with the receiver. Our study encompasses two distinct features: Pathloss and CIR-based PLA schemes, both developed for comparative assessment against our RIS-assisted schemes.
Pathloss-based PLA capitalizes on the free space pathloss (FSPL) model \cite{1697062}, while  CIR-based PLA utilizes direct channel CIR for authentication at the physical layer. For both baseline schemes, we employ binary hypothesis testing to represent the unique signatures of Alice and Eve. Furthermore, we gauge the performance of the test statistics in binary hypothesis testing employing error probabilities.



\subsection{Methodology}
We introduce two distinct methods for modeling and analyzing RIS-assisted PLA and non-RIS PLA systems. The MATLAB implementation of the system models is executed through \textit{analytical} and \textit{Monte-Carlo} simulations.

The latter uses random sampling techniques to model and analyze complex systems or processes. In contrast, analytical simulations rely on mathematical models and deterministic computations to derive precise solutions for specific problems. In our models, the analytical simulation uses the derived closed-form expressions of the PLA error probabilities, while the Monte-Carlo simulation constructs a comprehensive system where Alice and Eve engage in random transmissions to Bob. we set $10^8$ total number of transmissions for both transmitters which are randomly divided between the two according to uniform distribution. In a given time slot, we compute the outage events to record the error probabilities. 


\subsection{Results}

\subsubsection{Pathloss-based PLA}

In the pathloss-based PLA evaluation, we find the optimal phase shift to use in the RIS-assisted PLA simulations. We evaluate the models using PFA, PMD, and ROC in the two simulation methodologies, i.e., Monte-Carlo and analytical.

\begin{figure}[ht!]
    \centering
    \includegraphics[width=0.8\linewidth]{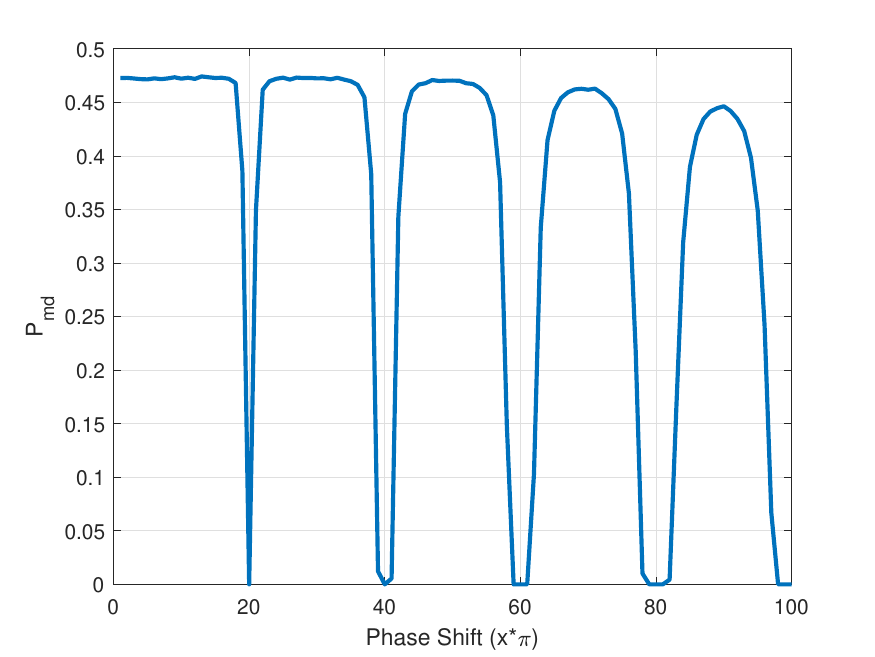}
    \caption{PMD against Phase shift in pathloss-based PLA}
    \label{fig:Pmd_v_PS_PL}
\end{figure}

\paragraph{RIS Phase Shift Design}
We discussed in Section \ref{s: phase shift design} the use of an exhaustive search method to find the optimal value of the phase shift that yields the minimal PMD value. Figure \ref{fig:Pmd_v_PS_PL} plots the curve between the PMD values and increasing phase shift gradient, showing that the curve exhibits regular zero crossings at fixed intervals, representing optimal phase gradient values when it reaches zero. It shows the efficacy of using RIS for PLA. It also represents how important it is to choose the phase shift of RIS in this scenario. Note that here we talk about a single phase-shift or single RIS element since pathloss of the RIS panel is almost the same as pathloss of a single RIS element. 

\paragraph{Probability of False Alarm (PFA)}

\begin{figure}[ht!]
\centering
  \begin{subfigure}[b]{0.8\columnwidth}
    \includegraphics[width=\linewidth]{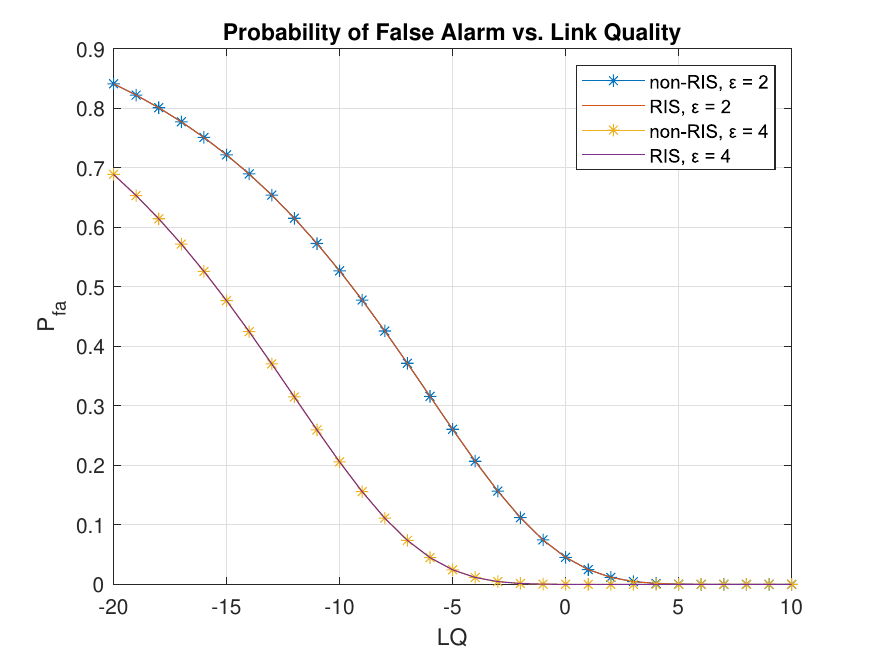}
    \caption{Analytical results}
    \label{fig:1}
  \end{subfigure}
   \hspace{1em} \\
  \begin{subfigure}[b]{0.8\columnwidth}
    \includegraphics[width=\linewidth]{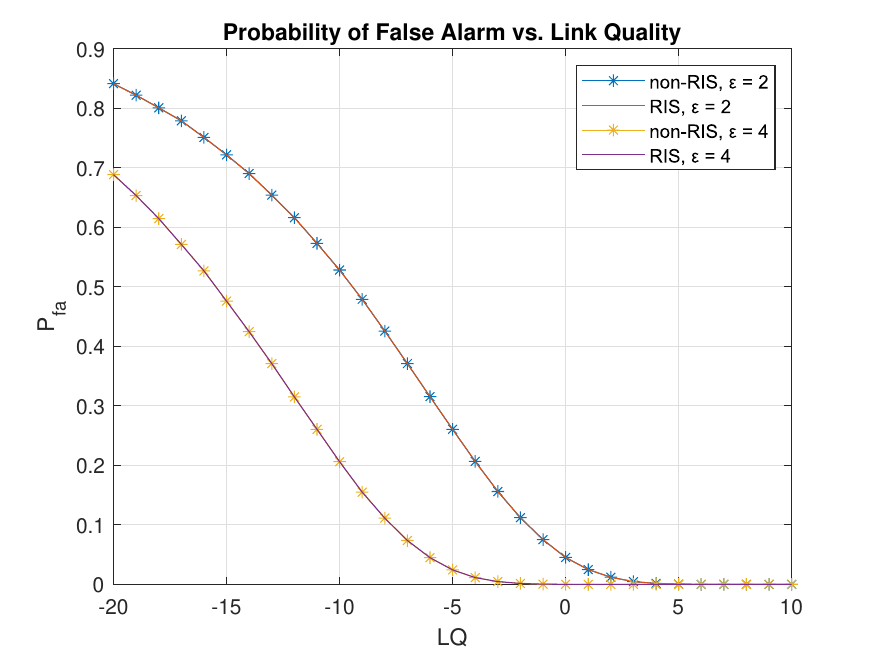}
    \caption{Monte-Carlo results}
    \label{fig:2}
  \end{subfigure}
    \caption{PFA against LQ [in dB] in pathloss-based PLA}
    \label{fig:PL_PFA}
\end{figure}



The curves presented in Figure \ref{fig:PL_PFA}, depict PFA curves in scenarios with and without RIS, comparing Monte-Carlo and analytical simulations results. Note that we plot Monte-Carlo results (upper subfigure of Figure \ref{fig:PL_PFA}) and analytical (lower subfigure of Figure \ref{fig:PL_PFA} ) separately for the sake of better exposition otherwise the nature and clutter of the curves would make it hard to read the figure. 
These curves exhibit a decreasing trend as the quality of the communication channel (the ratio of transmit to noise power) between transmitters and receivers improves. When the threshold increases, it influences both curves, resulting in smaller PFA values as the LQ improves. Notably, the curves for the RIS and non-RIS scenarios appear identical. This similarity arises from the characteristics of the PFA expressions, where the phase shift of the RIS does not affect the false alarm probability. One can also observe a perfect match between Monte-Carlo and analytical results which validates our analysis. 

\paragraph{Probability of Missed Detection (PMD)}
\begin{figure}[ht!]
\centering
  \begin{subfigure}[b]{0.8\columnwidth}
    \includegraphics[width=\linewidth]{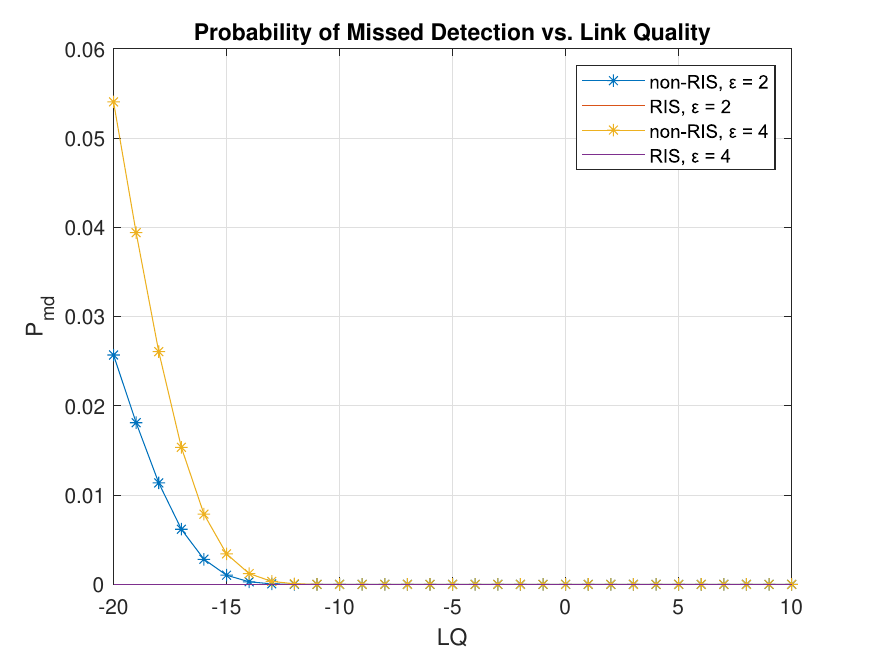}
    \caption{Analytical results}
    \label{fig:1}
  \end{subfigure}
   \hspace{1em} \\
  \begin{subfigure}[b]{0.8\columnwidth}
    \includegraphics[width=\linewidth]{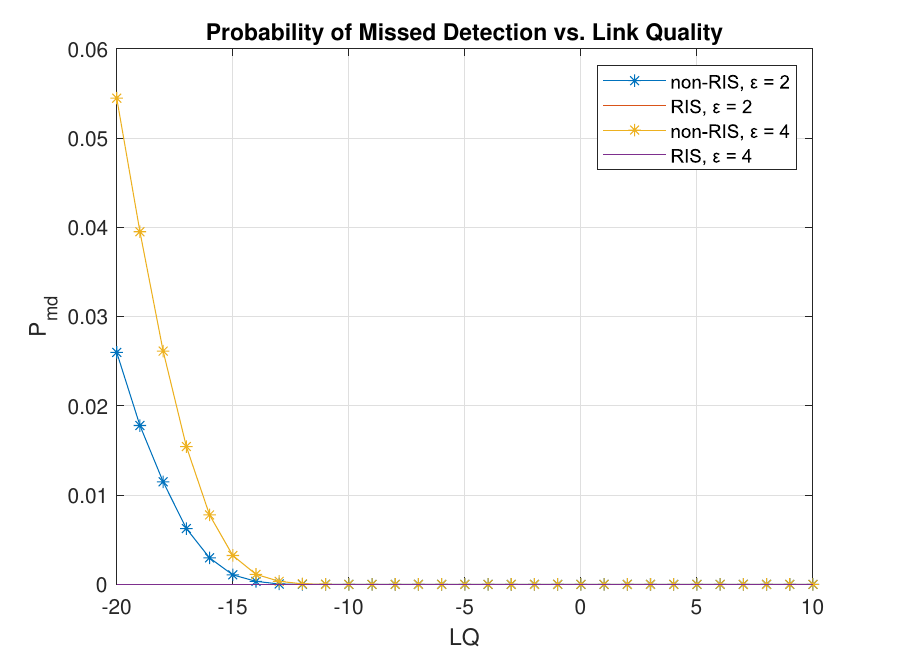}
    \caption{Monte-Carlo results}
    \label{fig:2}
  \end{subfigure}
    \caption{PMD vs LQ [in dB] in pathloss-based PLA}
    \label{fig:Pmd_PL_RIS_vs_NoRIS}
\end{figure}



As illustrated in Figure \ref{fig:Pmd_PL_RIS_vs_NoRIS} where in the absence of RIS, the PMD values consistently decrease with increasing LQ values. Furthermore, when we reduce the threshold, the PMD curves for the non-RIS scenario exhibits a further decrease. On the other hand, when configuring the RIS with the optimal phase shift obtained in Figure \ref{fig:Pmd_v_PS_PL}, we maintain zero PMD values for the RIS-assisted PLA model despite the varying LQ values or threshold values that show the important role of the RIS phase shift in the context of PLA.

\paragraph{Receiver Operating Characteristic (ROC)}
ROC curves serve as a summary metric to evaluate the performance of our authentication schemes. It illustrates the diagnostic ability of the proposed schemes, as its discrimination threshold $\epsilon$ varies in the range of 1 to 100 to provide a trade-off between the PFA and the probability of detection ($\text{P}_d=1-\text{P}_{md}$).

\begin{figure}[ht!]
\centering
  \begin{subfigure}[b]{0.8\columnwidth}
    \includegraphics[width=\linewidth]{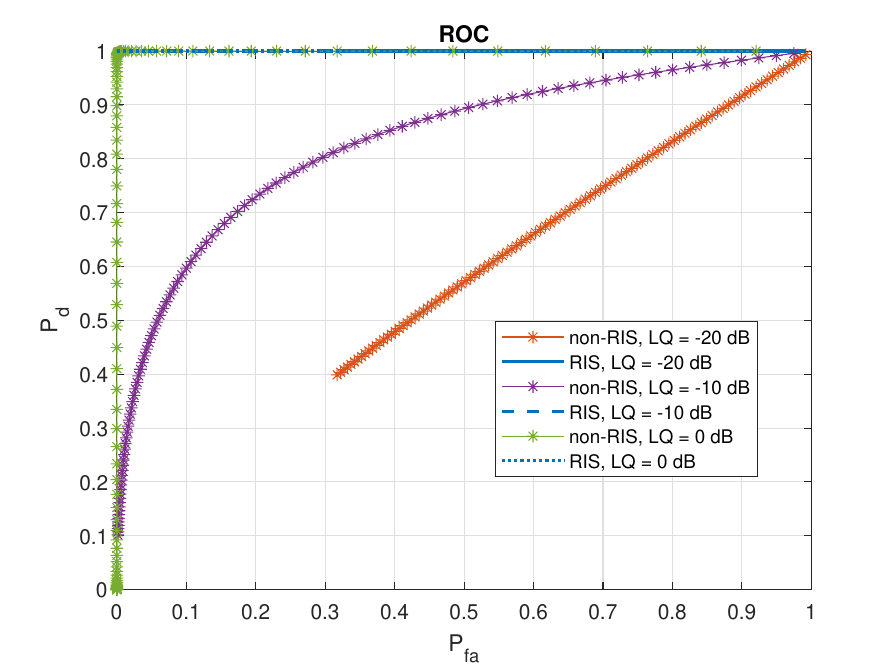}
    \caption{Analytical results}
    \label{fig:1}
  \end{subfigure}
   \hspace{1em} \\
  \begin{subfigure}[b]{0.8\columnwidth}
    \includegraphics[width=\linewidth]{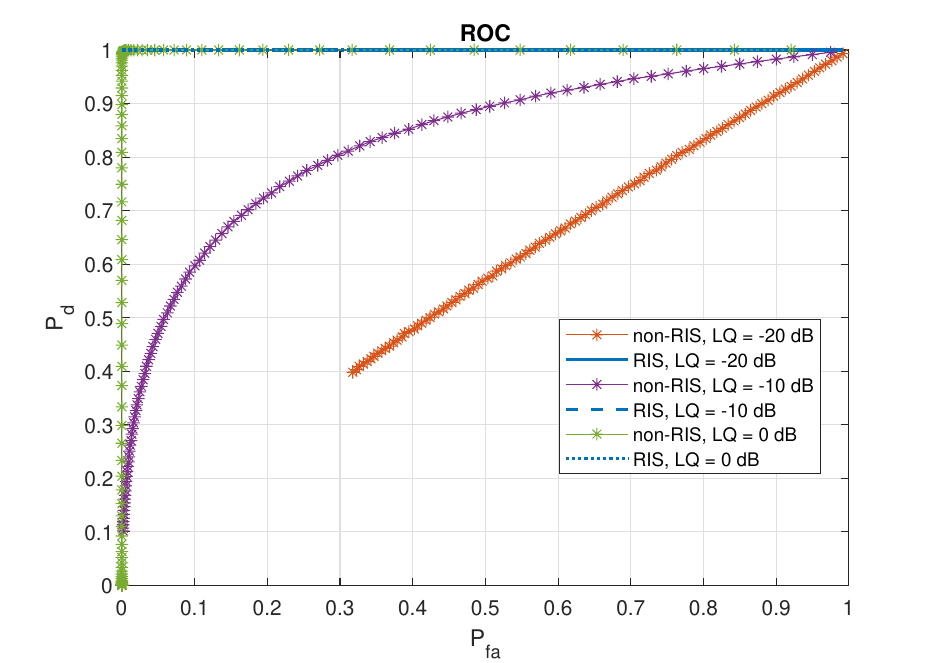}
    \caption{Monte-Carlo results}
    \label{fig:2}
  \end{subfigure}
    \caption{ROC Curves in Pathloss-based PLA}
    \label{fig:PL_ROC}
\end{figure}



In Figure \ref{fig:PL_ROC} , we plot the ($P_d = 1 - P_{md}$) against the $P_{fa}$ for RIS and non-RIS cases. With the absence of RIS, we can see that the system's performance in both simulation methods is enhanced by enhancing the LQ. However, we obtain the perfect ROC curve when integrating RIS into the system while using the optimal phase shift. The ROC curve passes through the top-left corner of the plot ($P_d$ = 1, $P_{fa}$ = 0) and proceeds horizontally, indicating high sensitivity and low false positive rate across all LQ values. Here, also the perfect match of the analytical and Mone-Carlo results validates the correctness of the analysis.

\subsubsection{CIR-based PLA}
 In the CIR approach, we estimate the channel gains as complex numbers characterized by \textit{magnitude} and \textit{phase} components. This section presents the results of PFA, PMD, and ROC by examining each of these fundamental components.

\begin{figure}[ht!]
\centering
  \begin{subfigure}[b]{0.45\columnwidth}
    \includegraphics[width=\linewidth]{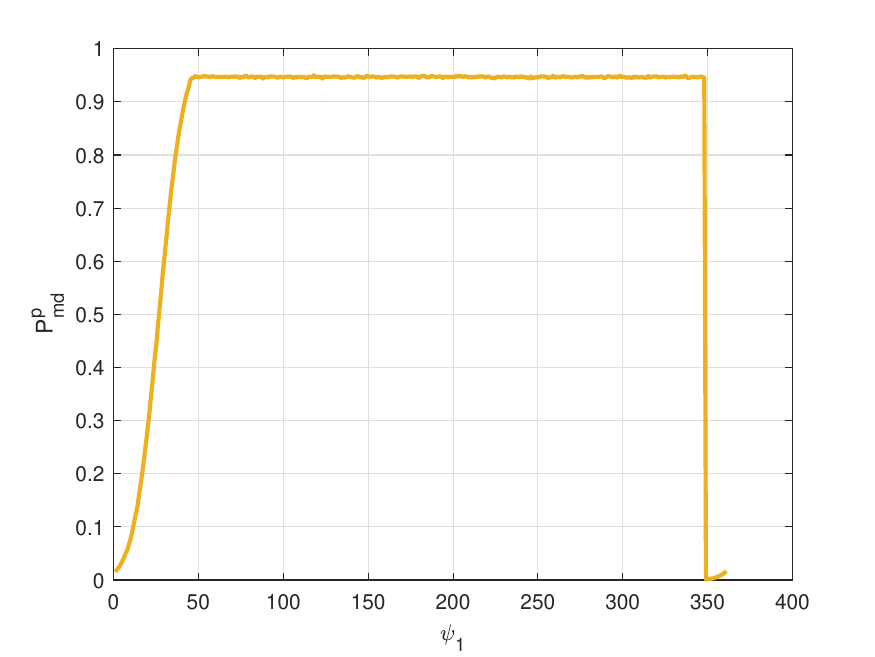}
    \caption{Phase shift of Element 1 in RIS configuration}
    \label{fig:1}
  \end{subfigure}
   \hspace{1em}
  \begin{subfigure}[b]{0.45\columnwidth}
    \includegraphics[width=\linewidth]{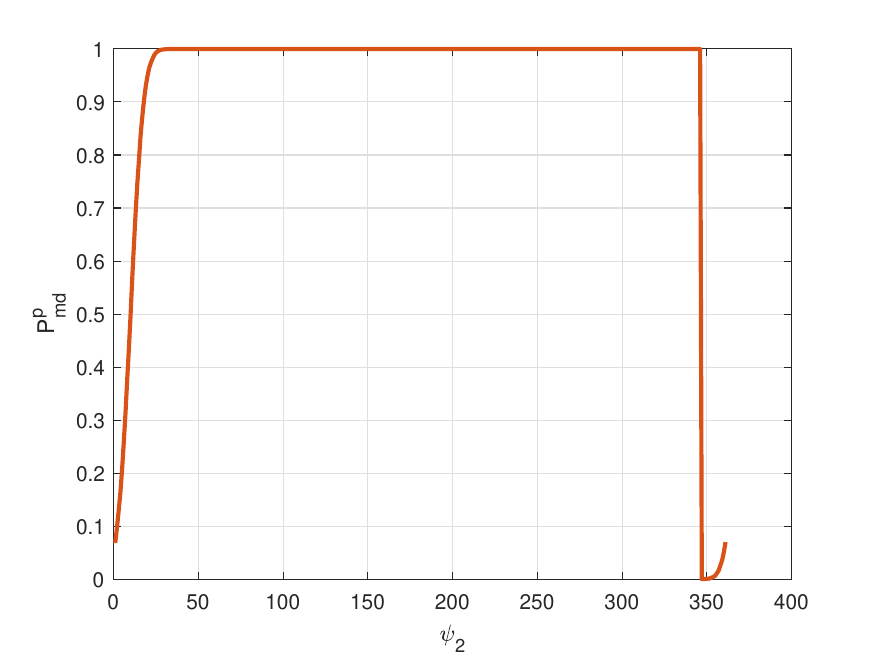}
    \caption{Phase shift of Element 2 in RIS configuration}
    \label{fig:2}
  \end{subfigure}
   \hspace{1em}
  \begin{subfigure}[b]{0.45\columnwidth}
    \includegraphics[width=\linewidth]{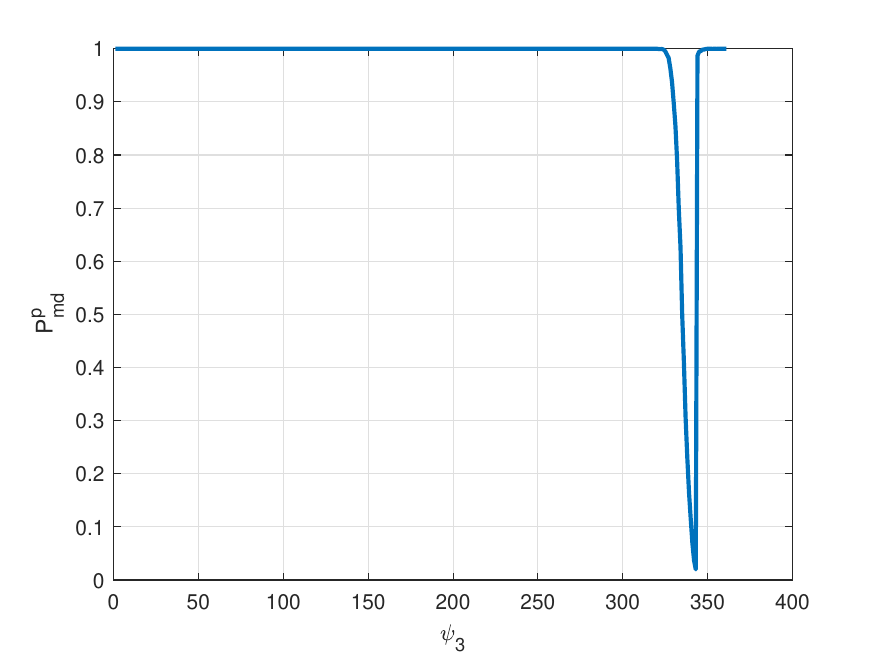}
    \caption{Phase shift of Element 3 in RIS configuration}
    \label{fig:2}
  \end{subfigure}
    \hspace{1em}
  \begin{subfigure}[b]{0.45\columnwidth}
    \includegraphics[width=\linewidth]{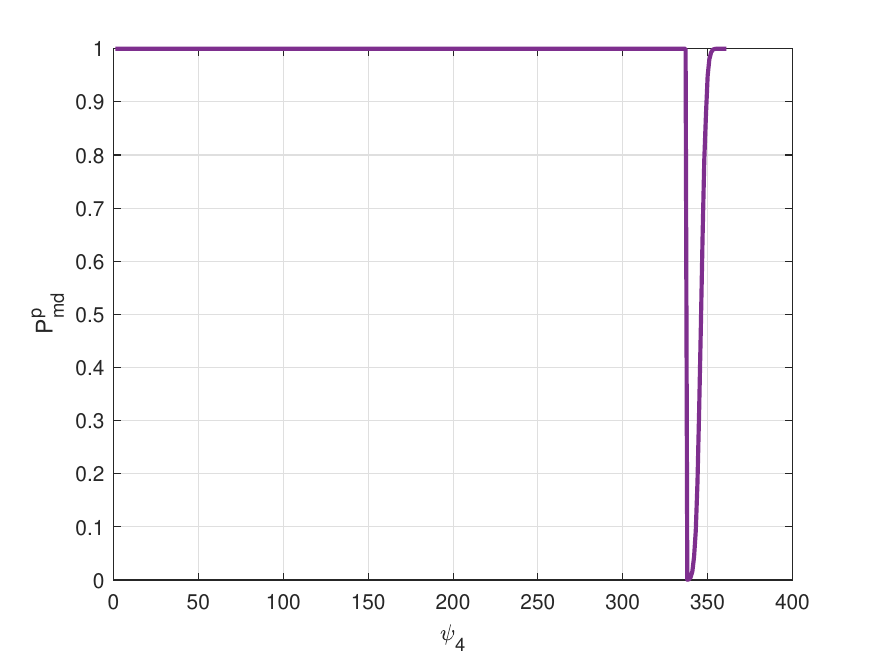}
    \caption{Phase shift of Element 4 in RIS configuration}
    \label{fig:2}
  \end{subfigure}
      \caption{PMD against Phase shift in CIR approach}
    \label{fig:Pmd_v_PS_CIR}
\end{figure}

\paragraph{RIS Phase Shift Design}
The expression in Section \ref{op-CIR} underscores the critical role of determining the RIS's elements' optimal phase shifts, which is crucial for minimizing the PMD values. Implementing the system setup specified in Table \ref{tab: pathloss parameters}, we integrate a 256-elements RIS into our model. We present four graphs in Figure \ref{fig:Pmd_v_PS_CIR} to visually depict the exhaustive search approach for identifying the optimal phase shift leading to a zero or minimal PMD value. These graphs illustrate the variation of PMD values across different phase settings $\psi \in [0,2\pi]$. Note that in CIR-based PLA, the main challenge is to do an exhaustive search as the computation is heavily dependent on the number of RIS  elements. Note that this figure is for the phase case, i.e, BHT$^p$ only as BHT$^m$ is not a function of phase shift of RIS. 

    
    
\paragraph{Probability of False Alarm (PFA)}

\begin{figure}[ht!]
\centering
  \begin{subfigure}[b]{0.8\columnwidth}
    \includegraphics[width=\linewidth]{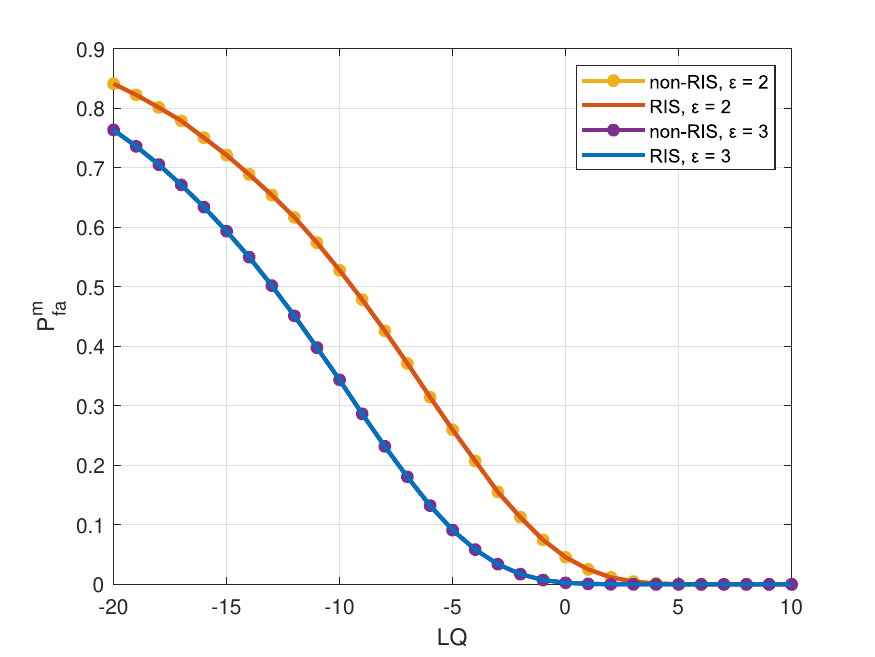}
    \caption{Analytical results}
    \label{fig:1}
  \end{subfigure}
   \hspace{1em} \\
  \begin{subfigure}[b]{0.8\columnwidth}
    \includegraphics[width=\linewidth]{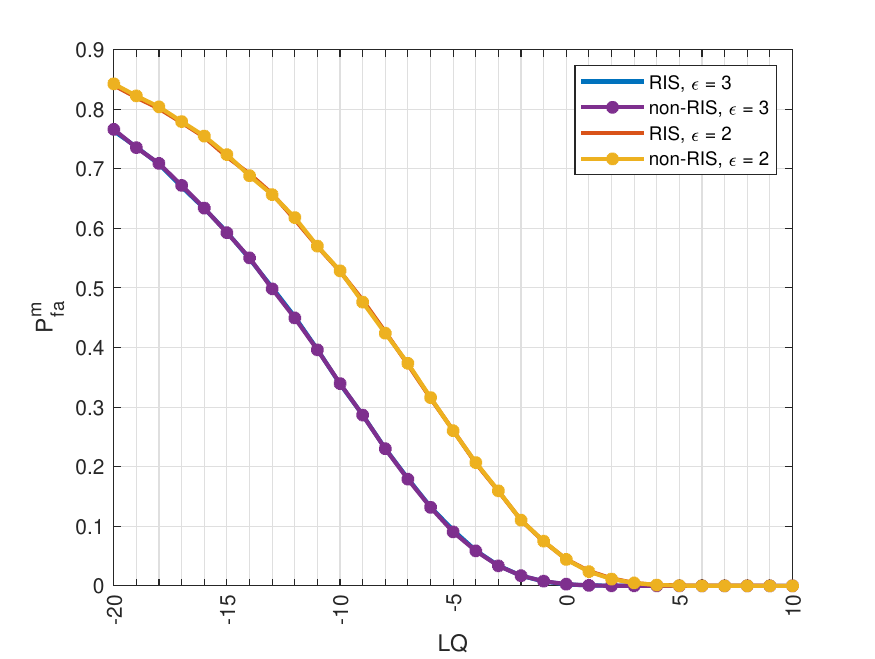}
    \caption{Monte-Carlo results}
    \label{fig:2}
  \end{subfigure}
    \caption{PFA for BHT$^m$ against LQ [in dB] in CIR-based PLA}
    \label{fig:Pfa_v_LQ_mag_CIR}
\end{figure}


\begin{figure}[ht!]
    \centering
    \includegraphics[width=0.8\linewidth]{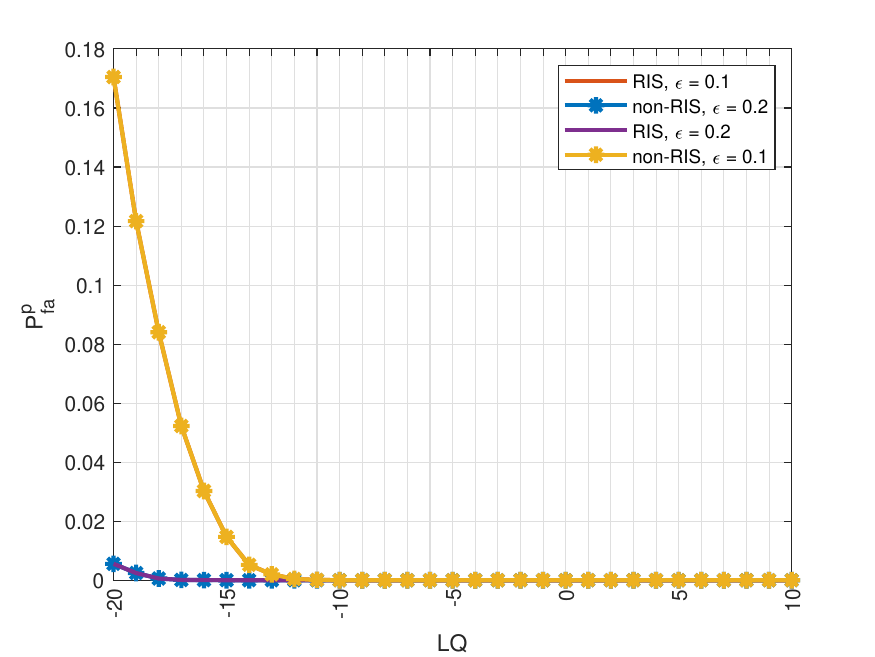}
    \caption{PFA for BHT$^p$ against LQ [in dB] in CIR-based PLA}
    \label{fig:Pfa_v_LQ_ph_CIR}
\end{figure}
The PFA graphs in Figures \ref{fig:Pfa_v_LQ_mag_CIR} and \ref{fig:Pfa_v_LQ_ph_CIR} show the magnitude and phase, respectively, of the PFA value as LQ increases. These graphs plot the behavior of the PFA value for both scenarios, with and without RIS, illustrating its response to changing LQ. When noise interference in the connection is minimized, one can see a drop in both $P^m_{fa}$ and $P^p_{fa}$ curves. The associated PFA values in both figures are further reduced by raising the threshold. Given the properties of the PFA expression, it remains invariant to the phase shift induced by RIS. As a result, RIS and non-RIS cases will exhibit identical curves in both graphs. In addition, the upper and lower subfigures in Figure \ref{fig:Pfa_v_LQ_mag_CIR} attest to the correctness of the analysis.

\paragraph{Probability of Missed Detection (PMD)}

\begin{figure}[ht!]
    \centering
    \includegraphics[width=0.8\linewidth]{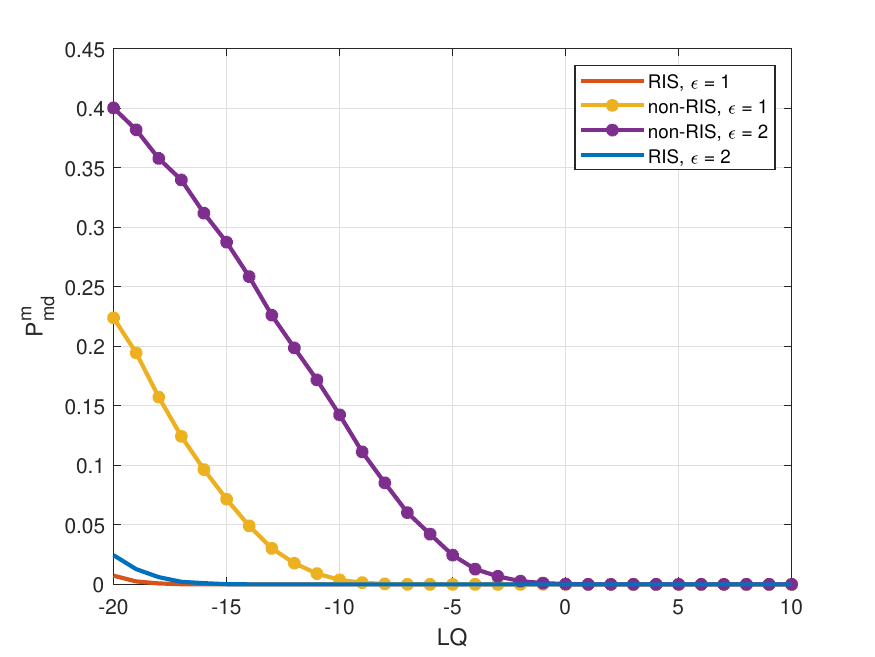}
    \caption{PMD for BHT$^m$ against LQ [in dB] in CIR approach}
    \label{fig:Pmd_v_LQ_mag_CIR}
\end{figure}

The graph presented in Figure \ref{fig:Pmd_v_LQ_mag_CIR} plots the PMD  values for BHT$^m$ in relation to the increase in LQ. The curve labeled ``non-RIS'' shows a decreasing pattern as LQ varies. Notably, a sharp decline in PMD values is observed when the threshold is decreased. However, the RIS curve also exhibits a decreasing trend with PMD values lower than those of the non-RIS curve. Interestingly, altering the threshold leads to only minor fluctuations in the PMD values for this curve.
\begin{figure}[ht!]
    \centering
    \includegraphics[width=0.8\linewidth]{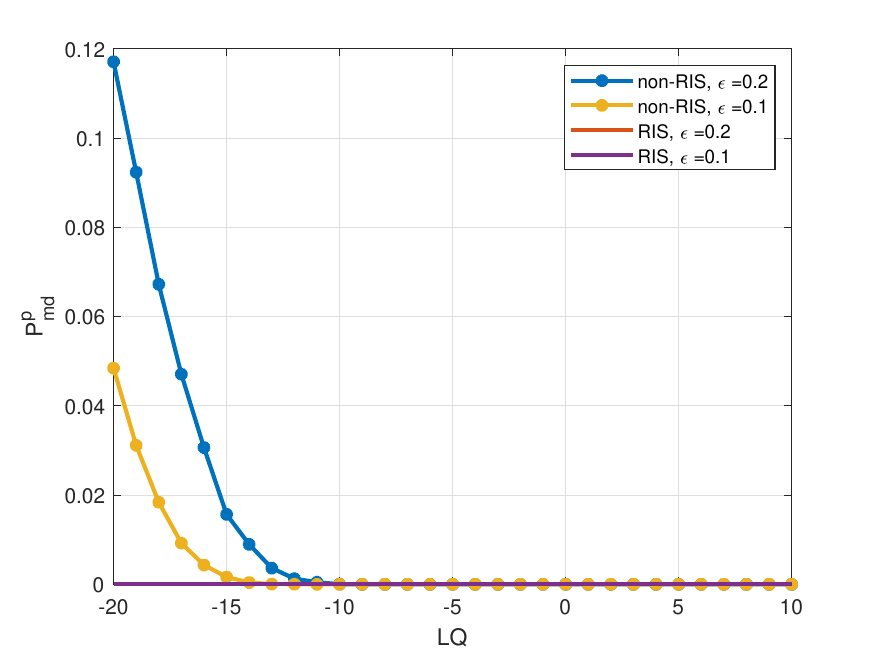}
    \caption{PMD (phase component) against LQ in CIR approach}
    \label{fig:Pmd_v_LQ_ph_CIR}
\end{figure}
Regarding the phase component of the PMD values, Figure \ref{fig:Pmd_v_LQ_ph_CIR} illustrates its correlation with increasing LQ. Similarly to Figure \ref{fig:Pmd_v_LQ_mag_CIR}, the PMD values for the non-RIS scenario decrease with increasing Link Quality, showing a further reduction as the threshold decreases. However, when configuring the RIS with an optimal phase shift, the PMD values in the RIS curve remain zero regardless of variations in threshold or LQ. 

\paragraph{Receiver Operating Characteristic (ROC)}
We generate ROC curves to evaluate a model's performance by illustrating the relationship between the PFA and the probability of detection $P_d^p$ across different threshold values.

\begin{figure}[ht!]
    \centering
    \includegraphics[width=0.8\linewidth]{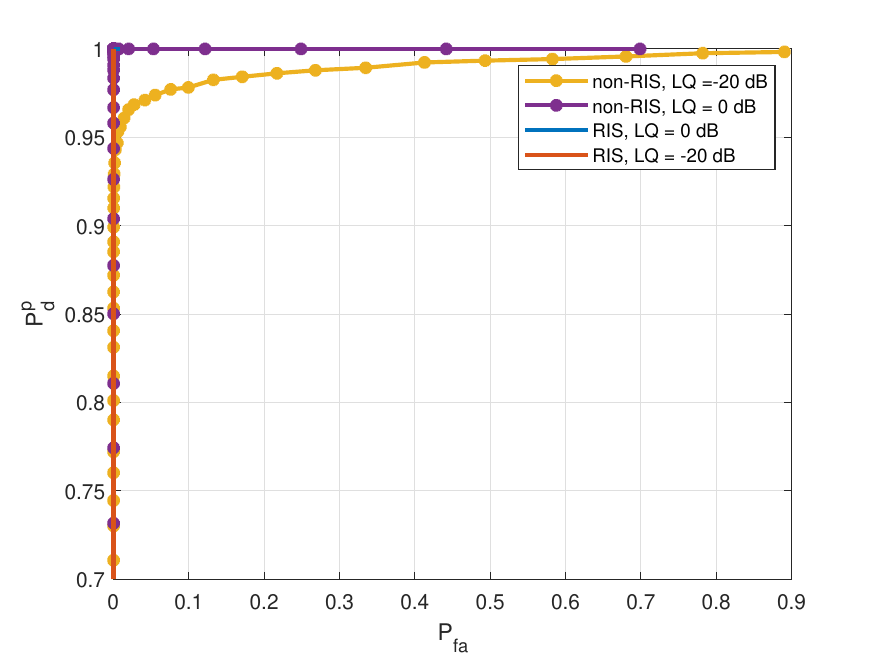}
    \caption{ROC curve in CIR approach}
    \label{fig:ROC_CIR_phase}
\end{figure}

In Figure \ref{fig:ROC_CIR_phase}, we plot an ROC curve, which illustrates the trade-off between the phase components of $P_d^p$ and $P_{fa}$. As LQ improves, the non-RIS curve approaches the ideal curve. In contrast, in the RIS case, the curve remains consistently optimal for all LQ values.

\subsection{Key lessons Learned}
In this subsection, we highlight the importance of this work by discussing the key lessons learned from this work.

\paragraph{Pathloss-based PLA}
In our proposed pathloss-based PLA, the performance of PFA and PMD metrics effectively demonstrates the favorable outcomes of our RIS-assisted PLA model. When contrasting the results between the RIS-assisted and non-RIS systems, it becomes evident that the PFA curves exhibit identical results. Therefore, relying solely on PFA rates may not suffice to assess the efficacy of RIS integration in the authentication process. Although reducing PFA rates is vital to avoid unnecessary disruptions within the system, PMD is of greater significance in ensuring the detection of potential threats. By optimizing the RIS phase shift, we successfully achieved a minimal PMD value, even amidst channel noise, effectively enhancing the robustness of our approach, as shown in the ROC curves.

The close resemblance between the results obtained from Monte-Carlo and analytical simulations in the Pathloss-based PLA indicates two points:
   \begin{itemize}
     \item \textbf{Validity of the Monte-Carlo method}, as it proves its efficacy in capturing the underlying mathematical relationships governing the system.
     \item \textbf{Robustness of analytical approach}, suggesting that the assumptions and simplifications inherent in the analytical model are reasonable approximations of reality.
   \end{itemize}

Even though the pathloss-based PLA shows promising results as an authentication factor in the physical layer, solely depending on it has its limitations. 
The fact that pathloss is a function of distance, introduces constraints to our PLA scheme. Considering the geometry of a circle,  if Alice's and Eve's positions are equidistant and on the same angle of incident to the norm of the RIS, Bob will not be able to distinguish the transmitter's fingerprint. Hence, the system would be vulnerable to threats.

\paragraph{CIR-based PLA}
Although the likelihood of encountering such a scenario is minimal, we recognize that it is not impossible, which requires further evaluation. Therefore, we used the CIR alongside realistic channel gains as a secondary approach. This will enable us to rigorously test our scheme using the same evaluation metrics, ensuring comprehensive assessment across various conditions. Given the complexity involved in deriving closed-form expressions for certain error probabilities within the CIR approach, we opt for numerical computation to obtain these probabilities.

The simulation results for the BHT$^m$ of PFA for CIR-based RIS-assisted and non-RIS cases exhibit similar behavior to the generated PFA graphs in the pathloss-based analysis. However, the phase component of the PFA graph illustrates even lower PFA values as LQ improves. Despite the impressive PFA values shown in the graphs, the impact of RIS on enhancing the CIR-based PLA is not apparent. In contrast, the PMD values demonstrate the significant impact of RIS on enhancing the PLA by substantially reducing the PMD values compared to the non-RIS case. With the utilization of the optimal RIS phase shift matrix, PMD phase values consistently remain at zero irrespective of variations in LQ and threshold settings. We see the comprehensive performance of our CIR-based PLA in the generated ROC, which shows a perfect curve when RIS is integrated into our model.

\section{Conclusion \& Future Work}
\label{s: s5}
In this paper, we proposed a novel Physical Layer Authentication (PLA) model for wireless communication systems, leveraging reconfigurable intelligent surface (RIS). Our approach focuses on integrating two key physical layer features: pathloss and channel impulse response (CIR), as foundational elements for our PLA schemes. Unlike existing models in the literature, we conducted a comprehensive comparative analysis to highlight the influence of RIS on PLA performance within wireless communication systems.  We introduced the innovative concept of RIS into the traditional system architecture, shaping our authentication scheme through binary hypothesis testing to discriminate between legitimate and malicious nodes. Using the main functionality of RIS in shifting the impingement signal to the desired phase, we were able to demonstrate the effectiveness of RIS in enhancing PLA compared to its non-RIS counterpart.


The proposed authentication model could be further extended and incorporate several enhancements to boost its efficacy:
\begin{itemize}
    \item Incorporating Machine Learning (ML) models for phase shift optimization enhances efficiency by navigating possibilities more swiftly, avoiding the computational burdens of exhaustive searches. ML's capacity to rapidly learn from data not only streamlines optimization but also facilitates adaptation, outperforming exhaustive searching in speed and scalability.
    \item Exploring the integration of diverse physical layer features, or their combination, holds promise for enhancing authentication accuracy and broadening the model's compatibility across diverse systems.
    \item Conducting an analytical simulation for the CIR-based PLA model in RIS and non-RIS systems to ensure robustness of the model, by deriving the closed-form expressions for the performance metrics.
\end{itemize}
\footnotesize{
\bibliographystyle{IEEEtran}
\bibliography{references}
}

\vfill\break

\end{document}